\documentclass{kapproc} 
\usepackage{psfig}

\let\footnote\savefootnote

\let\footnotetext\savefootnotetext 

\let\footnoterule\savefootnoterule 

\kluwerbib

\begin{document}

\articletitle{Binary Population Synthesis: Low- and Intermediate-Mass
X-Ray Binaries}


\author{Ph.\ Podsiadlowski}
\affil{Oxford University\\
OX1 3RH, U. K.}
\email{podsi@astro.ox.ac.uk}

\author{S. Rappaport, E. Pfahl}
\affil{Department of Physics,\\ 
Massachusetts Institute of Technology,\\
Cambridge, MA 02139}
\email{sar@mit.edu, pfahl@mit.edu}

\begin{abstract}
As has only recently been recognized, X-ray binaries with
intermediate-mass secondaries are much more important than previously
believed.  To assess the relative importance of low- and
intermediate-mass X-ray binaries (LMXBs and IMXBs), we have initiated
a systematic study of these systems consisting of two parts: an
exploration of the evolution of LMXBs and IMXBs for a wide
range of initial masses and orbital periods using detailed binary
stellar evolution calculations, and an integration of these results
into a Monte-Carlo binary population synthesis code. Here we present
some of the main results of our binary calculations and some
preliminary results of the population synthesis study for a
``standard'' reference model. While the inclusion of IMXBs improves
the agreement with the observed properties of ``LMXBs'', several
significant discrepancies remain, which suggests that additional
physical processes need to be included in the model.\footnote{To 
appear in Podsiadlowski et al.\ 2001,
in, D. Vanbeveren, ed., {\it The Influence of Binaries on Stellar
Population Studies} (Kluwer, Dordrecht), p. 355.}

\end{abstract}


\section{Introduction}
Until recently it was generally believed that X-ray binaries fall
into two distinct classes, low-mass X-ray binaries (LMXBs) and high-mass
X-ray binaries (HMXBs), with Her X-1 being the only exception that does not
really belong to either class (for a general review see, e.g., 
Bhattacharya \& van den Heuvel 1991). However, in the systems classified
as ``LMXBs'' the nature of the secondary is usually unknown. Whenever it has 
been possible in recent years  to determine the nature
of the secondaries in so-called ``LMXBs'', these did not actually comply 
with the
standard paradigm for LMXBs. First, van Kerkwijk et al.\ (1992) showed that the
X-ray binary Cyg X-3, until then generally considered a ``proto-type
LMXB'', actually contained a Wolf-Rayet star and hence, if anything, should 
be classified as a HMXB. More recently, it was recognized that the 
second-brightest X-ray source in the same constellation, Cyg X-2, also does
not belong to the group of LMXBs, but is the descendant of
an intermediate-mass X-ray binary (IMXB; see Sect.\ 2.1). This latter
discovery immediately raises the question whether many other, perhaps even 
most systems classified as ``LMXBs'' are in actual fact ``IMXBs''. 
\par
In order to determine the relative importance of LMXBs and IMXBs we have
initiated a systematic binary population synthesis study
(Pfahl, Rappaport \& Podsiadlowski 2001; Podsiadlowski, Rappaport \&
Pfahl 2001). Our approach is different from most previous studies in so far as
we use realistic binary evolution calculations instead of simplified
analytical prescriptions to model individual binary sequences.\par
In Section~2 of this contribution we first discuss the developments that
have led to the re-assessment of the importance of IMXBs;
in Section~3 we describe our comprehensive set
of binary calculations highlighting some of the main results
of these calculations, and in Section~4 we show how we are implementing these
sequences into a binary population synthesis code and illustrate this 
with some preliminary results.
For a discussion of the related problem, the formation of X-ray 
binaries and millisecond pulsars in globular clusters, we refer
to the contribution by Rasio (Rasio 2001).

\section{The Importance of Intermediate-\\ Mass X-Ray Binaries (IMXBs)}

Until recently IMXBs had received relatively little attention
(see, however, Pylyser \& Savonije 1988, 1989 and in the context of black-hole
transients Kolb 1998; Reg\"os et al.\ 1998). One of the main reasons for
this neglect is the fact that only one such system, Her X-1, had 
unambiguously been identified in the past. This changed recently with two new
developments. First, Davies \& Hansen (1998) studying dynamical interactions
in globular clusters found that IMXBs are much easier to form dynamically
than LMXBs and speculated that these IMXBs, which do not exist in globular
clusters at the present epoch, might be the progenitors of the observed 
millisecond pulsars rather than the presently observed LMXBs.

The second development, which spurred our own investigation, was a
re-assessment of the evolutionary status of the X-ray binary
Cyg X-2.

\subsection{The case of Cygnus X-2}

\begin{figure}[t]
\hfill\hbox{\psfig{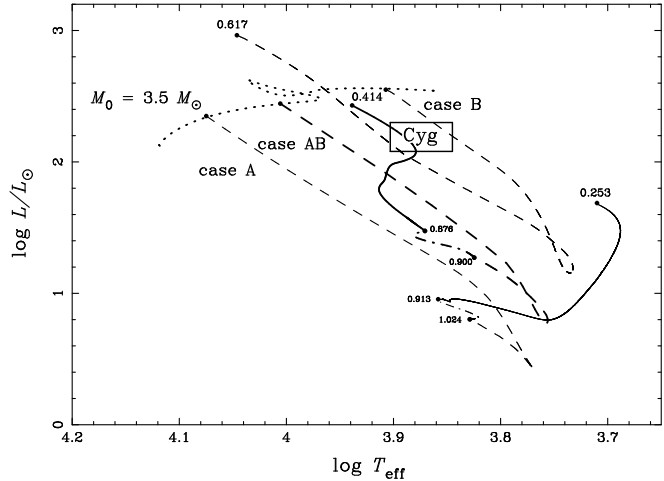}}\hfill
\caption {Evolutionary tracks of the secondaries in
three binary calculations in the Hertzsprung-Russell diagram.  The
secondary has an initial mass of $3.5\,M_{\odot}$ and the primary, assumed to
be a neutron star, an initial mass of $1.4\,M_{\odot}$ in all calculations. The
dotted curve shows the track of a $3.5\,M_{\odot}$ star without mass
loss. The mass-loss tracks, labelled case A, AB and B, start at
different evolutionary phases of  the secondary (`case A': the middle 
of the main sequence; `case AB': the end of the main sequence; 
`case B': just after the main-sequence). The dashed portions in each track
indicate the rapid initial mass-transfer phase, the dot-dashed and
dashed portions  the slow phases where mass-transfer is driven by hydrogen 
core burning and hydrogen shell burning, respectively (only in case A 
and case AB). The beginning and end points of the various phases are
marked by solid bullets, the small figures next to them give the
mass of the secondary at these points. The boxed region labelled `Cyg'
indicates the observationally determined parameter region for the secondary
in Cyg X-2. The tracks after mass transfer has ceased are not shown (from
Podsiadlowski \& Rappaport 2000).}
\end{figure}

The spectroscopic observations of Cyg X-2 by Casares, Charles \& 
Kuulkers (1998) combined with the modelling of the ellipsoidal light 
curve (Orosz \& Kuulkers 1999) showed that the secondary in Cyg X-2 was 
indeed a low-mass star with $M_2=0.6\,\pm\,0.13\,M_{\odot}$. However, the same
observations also showed that the spectral type of the companion is,
to within two subclasses, A9III and that {\em the spectral type does not 
vary with orbital phase.} Since this almost certainly constitutes the intrinsic
spectral type of the secondary, it means that the secondary is far
too hot and almost a factor of 10 too luminous to be consistent with
a low-mass subgiant with an orbital period of 9.84 days (see Podsiadlowski
\& Rappaport 2000).\par
A resolution of this paradox was found independently by King \& Ritter
(1999) and Podsiadlowski \& Rappaport (2000) (also see Kolb et al.\ 2000;
Tauris, van den Heuvel, \& Savonije 2000) who showed that the 
characteristics of Cyg X-2
can be best understood if the system was the descendant
of an IMXB where the secondary initially had a mass of $\simeq 3.5\,M_{\odot}$
and lost most of its mass in very non-conservative case AB or case B
mass transfer. This is illustrated in Figures~1 and 2 (from
Podsiadlowski \& Rappaport 2000).

\begin{figure}[t]
\hfill\hbox{\psfig{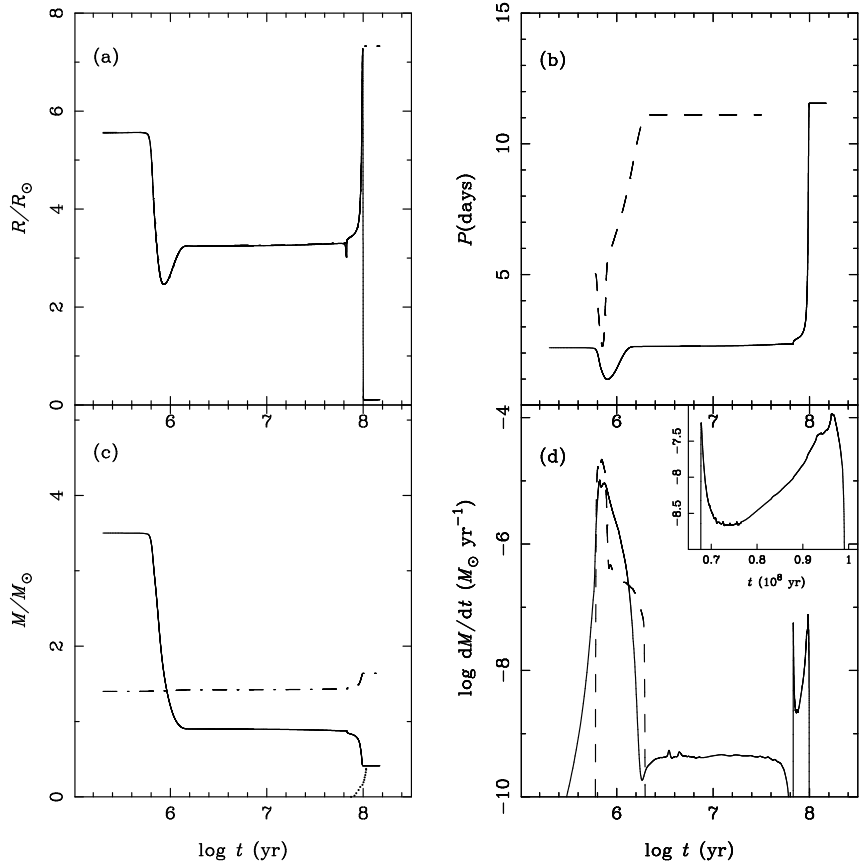}}\hfill
\caption{Key binary parameters for the case AB binary
calculation as a function of time (with arbitrary offset). Panel (a):
radius (solid curve) and Roche-lobe radius (dot-dashed curve) of the
secondary; panel (b): the orbital period (solid curve); panel (c):
the mass of the secondary (solid curve) and the primary (dot-dashed
curve); panel (d): the mass-loss rate from the secondary (solid
curve); the inset shows a blow-up of the second slow mass-transfer phase
(hydrogen shell burning). The dashed curves in panel (b) and (d)
show the orbital period and mass-transfer rate for the case B calculation
for comparison (from Podsiadlowski \& Rappaport 2000).}
\end{figure}

Thus, Cyg X-2 provides observational proof that IMXBs can eject most
of the mass that is being transferred from the secondary (perhaps
in a way similar to what happens in SS433) and subsequently resemble
classical LMXBs. This raises the immediate question of what fraction
of so-called ``LMXBs'' are in reality IMXBs or their descendants.
It is also worth noting that IMXBs are much easier to form than LMXBs
since they do not require the same amount of fine-tuning as LMXBs
(see Bhattacharya \& van den Heuvel 1991). Furthermore, if there is a large
fraction of IMXBs this will not only affect the period and the 
luminosity distribution of ``LMXBs'', but also has important implications 
for the so-called ``LMXB'' -- ms pulsar birthrate problem (e.g.,
Kulkarni \& Narayan 1988), all issues that can only be properly addressed
with the help of binary population synthesis.

\section{Binary Sequences for LMXBs and IMXBs}

In order to systematically investigate LMXBs and IMXBs, we performed
a series of $\sim$100 binary evolution calculations, covering the
mass range of 0.8 to $7\,M_{\odot}$ and all evolutionary phases from 
early case A to late case B mass transfer.

\subsection{Model assumptions}

All calculations were carried out with an up-to-date, standard Henyey-type
stellar evolution code (Kippenhahn, Weigert, \& Hofmeister 1967),
which uses OPAL opacities (Rogers \& Iglesias 1992) complemented with
those from Alexander \& Ferguson (1994). We use solar metallicity
($Z=0.02$), a mixing-length parameter $\alpha=2$ and assume 0.25
pressure scale heights of convective overshooting from the core
(Schr\"oder, Pols, \& Eggleton 1997). The mass-transfer rate, $\dot{M}$
is calculated explicitly using the formalism of Ritter (1988). 
We further assume that half the mass transferred is accreted by
the neutron star up to the Eddington limit, taken to be $2\times 10^{-8}\,
M_{\odot}\,\hbox{yr}^{-1}$. Any matter leaving the system carries
with it the specific angular momentum of the accretor. Furthermore,
we include angular-momentum losses due to gravitational radiation and
magnetic braking (for stars with convective envelopes), for the latter 
using the formalism of Verbunt \& Zwaan (1981).

\subsection{Selected results}

\begin{figure}[t]
\hfill\hbox{\psfig{figure=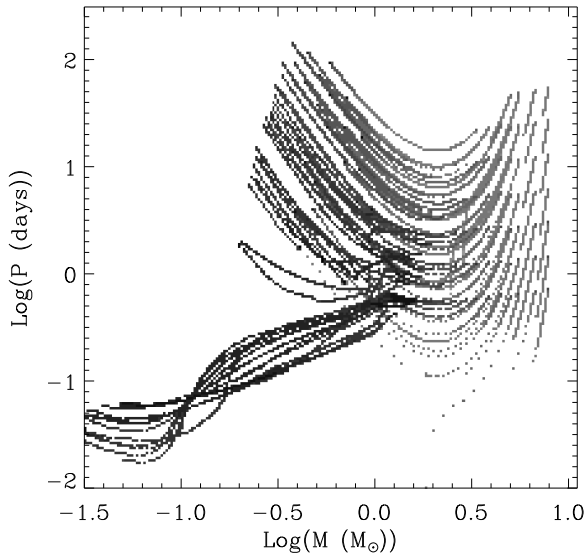,height=3.3truein,angle=0}}\hfill
\caption{Binary evolution tracks in the $\log M_2\,$--$\,\log P_{\rm orb}$
plane. The degree of shading measures the amount of time spent
in a particular part of the diagram.}
\end{figure}

Figure~3 presents the evolutionary tracks of our binary sequences in
the secondary-mass orbital-period ($\log M_2\,$--$\,\log P_{\rm orb}$)
plane. Broadly the sequences can be divided into three classes: (1)
and (2) systems evolving to long periods and short periods,
respectively, separated by the well-known bifurcation period of just
under 1 day, and (3) more massive systems experiencing dynamical mass
transfer and spiral-in (more easily discerned in Fig.~4). What this
figure does not show, however, is the actual variety in these
sequences. Some 70 of the $\sim$100 sequences are qualitatively
different with respect to the importance and the order of different
mass-transfer driving mechanisms, the occurrence of detached phases,
the final end products, etc. In this context it is worth noting that
there are very few sequences that resemble the classical
cataclysmic-variable (CV) evolution where mass transfer is solely
driven by gravitational radiation and magnetic braking. This may be
important in explaining some of the fundamental differences observed
between CVs and LMXBs/IMXBs.

\subsubsection{Delayed dynamical instability}

The majority of systems in which the initial secondary mass is 
$4\,M_{\odot}$ and all systems more massive than $4.5\,M_{\odot}$ experience 
dynamical mass transfer, resulting in the spiral-in of the neutron star inside
the secondary. The final outcome of this channel is very uncertain at present,
but it may be a  spun-up neutron star or a black hole, single or in
a very compact binary. In all systems, where the secondary is still on
the main sequence when mass transfer starts, this dynamical instability
is delayed (see Hjellming \& Webbink 1987) since the secondaries initially
have radiative envelopes which are stable against dynamical mass loss.
Dynamical instability occurs once the radiative part of the envelope
with a rising entropy profile has been lost and the core with a relatively
flat entropy profile starts to determine the reaction of the star to mass
loss. This delay may last for up to $10^6\,$yr; during this time the system
should still be detectable as a X-ray binary, with a very high mass-transfer
rate and quite possibly some unusual properties (such as SS433?) in the last 
$10^4\,$--$\,10^5\,$yr before the onset of the dynamical instability.

\subsubsection{Evolutionary end products}

\begin{figure}[t]
\hfill\hbox{\psfig{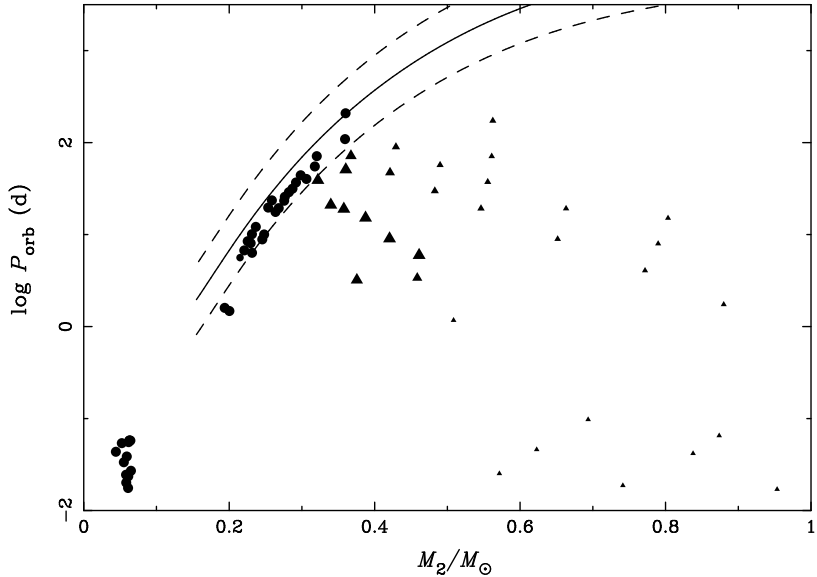}}\hfill
\caption{The end points of the binary sequences in the $M_2\,$--$\,\log P_{\rm
orb}$ plane.  Circles and
triangles indicate that the secondaries are He white dwarfs and HeCO
white dwarfs, respectively. The size of the symbols indicates how much
mass the neutron stars have accreted (systems with large symbols may
be reasonably expected to contain millisecond pulsars).  The low-mass,
ultracompact systems ($M_{2} < 0.1\,M_{\odot}$, $P_{\rm orb} <
0.1\,$d) are plotted when the systems pass through the orbital-period
minimum.  The more massive, ultracompact systems have experienced
dynamical mass transfer and a spiral-in phase and are plotted at the
orbital period at which a common envelope can be ejected energetically
(although in reality most of these systems are likely to merge
completely). The solid and dashed curves give the range of
the white-dwarf mass -- orbital-period relation for wide binary
radio pulsars (from Rappaport et al.\ 1995).}
\end{figure}

In Figure~4 we show the final position of the majority of the sequences
calculated in the $M_2\,$--$\,\log P_{\rm orb}$ plane,
where the size of the symbols indicates how much mass the neutron stars
have accreted (systems with large symbols may be reasonably expected
to contain millisecond pulsars). Circles indicate that the secondary
ends its evolution as a He white dwarf. Note that all systems with
$P_{\rm orb} > 0.1\,$d lie very close
to the relation between white-dwarf mass and orbital period for
wide binary radio pulsars calculated by Rappaport at al.\ 1995 (solid
and dashed curves). In systems with triangle symbols, the secondaries 
burn helium in the core in a hot OB subdwarf phase
(after mass transfer has been completed) and become HeCO white
dwarfs (i.e., non-standard white dwarfs with a CO core and a large
He envelope). Note that these HeCO white dwarfs can have a mass as low
as $0.3\,M_{\odot}$, the minimum mass for helium ignition in non-degenerate
cores (see, e.g., Kippenhahn \& Weigert 1990). Most of these systems
lie well below the white-dwarf -- orbital period relation without having
experienced a common-envelope phase (also see Podsiadlowski \& Rappaport
2000; Tauris et al.\ 2000). It is not immediately clear whether the 
fact that many of these relatively low-mass white dwarfs have CO cores 
has detectable, observational consequences.

Finally, it is worth noting that most of the white dwarfs with masses
$< 0.6\,M_{\odot}$ experience several dramatic hydrogen shell flashes
(typically 2 to 4) before settling on the white-dwarf cooling sequence
(these are reminiscent of the final helium shell flashes for relatively
massive post-AGB stars; Iben et al.\ 1983). During these flashes,
the luminosity typically rises by a factor of 1000 and the radius increases
by a factor of 10 or more on timescales of a few decades. Indeed, during
these flashes the secondaries tend to fill their Roche lobes again, leading
to several short mass-transfer phases with very high mass-transfer rates
driven entirely by the dynamics of these flashes.

\subsection{The formation of\\ ultra-compact binaries}

\begin{figure}[t]
\hfill\hbox{\psfig{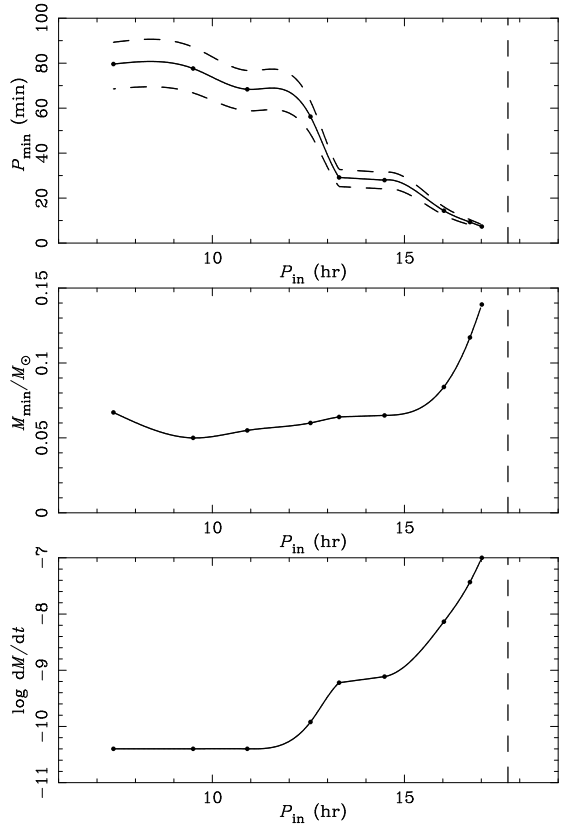}}\hfill
\caption{The formation of ultra-compact LMXBs. {\em Top Panel:\/} The
minimum orbital period versus initial orbital period. {\em Middle
and Bottom Panels:\/} The mass of the secondary and the mass-transfer
rate, respectively, at the minimum period. All calculations start
with a $1.4\,M_{\odot}$ neutron star and a $1\,M_{\odot}$ secondary.
The vertical dashed line indicates the initial orbital period above which 
systems become wider rather than more compact (i.e., the bifurcation period).
The dots indicate the results of the calculated sequences. The dashed
curves in the top panel indicate the range of minimum periods if mass
transfer is assumed to be either fully conservative (upper curve) or fully 
non-conservative (lower curve).}
\end{figure}

As Figures~3 and 4 show, systems with initial orbital periods somewhat
less than $\sim 1\,$d lead to ultra-compact binaries with minimum
orbital periods in the range of 20$\,$--$\,90\,$min. The shortest
period is not much longer than the $11\,$min period in the X-ray
binary 4U 1820-30 in the globular cluster NGC 6624 (Stella et al.\
1987). Unlike the two popular models for the formation of this system,
this evolutionary channel involves neither a direct collision (Verbunt
1987) nor a common-envelope phase (Bailyn \& Grindlay 1987; Rasio,
Pfahl, \& Rappaport 2000) and therefore constitutes an attractive
alternative scenario for 4U 1820-30 (this model was first suggested by
Tutukov et al.\ 1987 and was examined in some detail by Fedorova \&
Ergma 1989). \par

To determine the shortest orbital period that can be attained through this
channel we performed a separate series of binary calculations for
a $1\,M_{\odot}$ secondary (as would be appropriate for the initial mass
of the secondary in 4U 1820-30). We found,
confirming the earlier results of Fedorova \& Ergma (1987), that, if the 
secondaries start mass transfer near the end of core hydrogen burning (or, in
fact, just beyond), the secondaries transform themselves into
degenerate helium stars and that orbital periods as short as 
$5\,$min can be attained without the spiral-in of the neutron star 
inside a common envelope.\par

The results of this exploration are summarized in Figure~5. As the top
panel shows, there is a fairly large range of initial orbital periods 
($13\,$--$\,18\,$hr) which leads to ultra-compact LMXBs with a
minimum orbital period less than $30\,$min. Indeed, it is remarkable
that one of our binary sequences, with an initial orbital
period around $17\,$hr, appears to be a suitable sequence to explain
all five LMXBs in globular clusters whose orbital periods are presently
known, from the system with the longest period (AC211/X2127+119 in
M15; $P_{\rm orb}=17.1\,$hr; Ilovaisky et al.\ 1993) to the 11-min binary. 
Furthermore, systems with an initial period in the range of $13\,$--$\,18\,$hr 
are quite naturally produced as a result of the tidal capture of a neutron
star by a main-sequence star (Fabian, Pringle, \& Rees 1975) and are not
the generally expected outcome of a 3- or 4-body exchange 
interaction (see, e.g., Rasio et al.\ 2000; Rasio 2001). This
suggests to us that it may be premature to rule out tidal capture as a formation
scenario for LMXBs in globular clusters as has often been done in recent
years.

\section{Binary Population Synthesis}

        For our population synthesis study, we determine, starting from
primordial binaries, the mass of the secondary and the orbital period at
the beginning of the X-ray binary phase.  We then choose the binary
sequence from our two-dimensional grid that is closest to this point in the
($M_{2},P_{\rm orb}$) plane to represent the X-ray binary phase.  As our
systematic exploration of LMXB/IMXB evolution has shown, there is such a
large variety of possible binary sequences that one cannot hope that all of
them can be well represented by simple (semi-)analytic prescriptions ---
as is often done in such studies.  We have therefore chosen this improved
approach, wherein we directly integrate our grid of realistic binary
sequences into the Monte Carlo code.  Apart from this, most of our model
assumptions are fairly standard.

\subsection{Model assumptions}

\begin{figure}[t]
\hfill\hbox{\psfig{figure=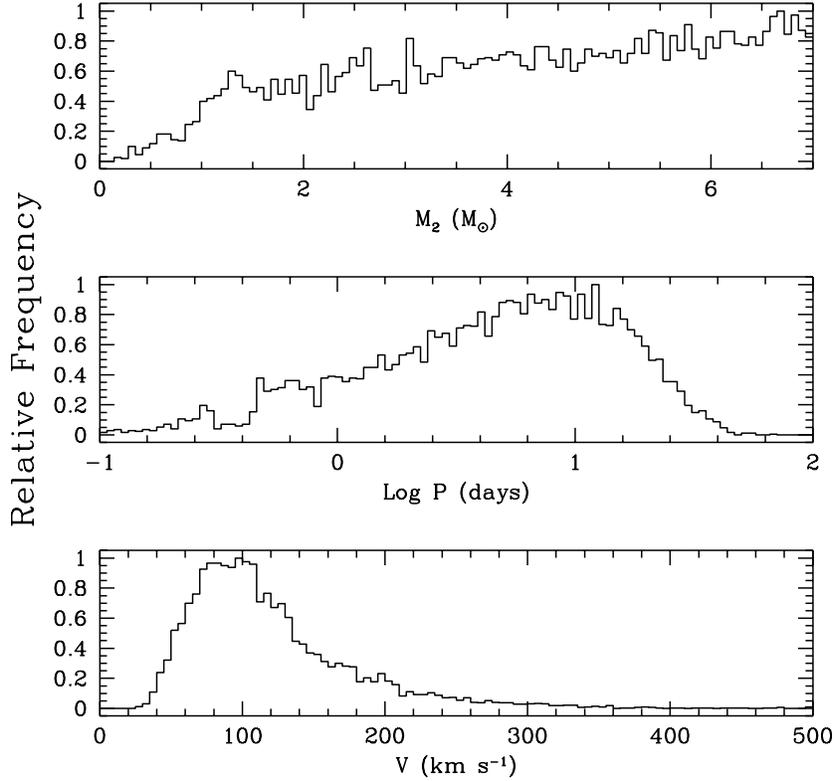,height=4.45truein,angle=0}}\hfill
\caption{The distributions of secondary mass (top panel), orbital
period (middle panel) and system velocity (bottom panel) at the
beginning of the mass-transfer phase onto the neutron star for our 
standard reference model.}
\end{figure}

\begin{figure}[t]
\hfill\hbox{\psfig{figure=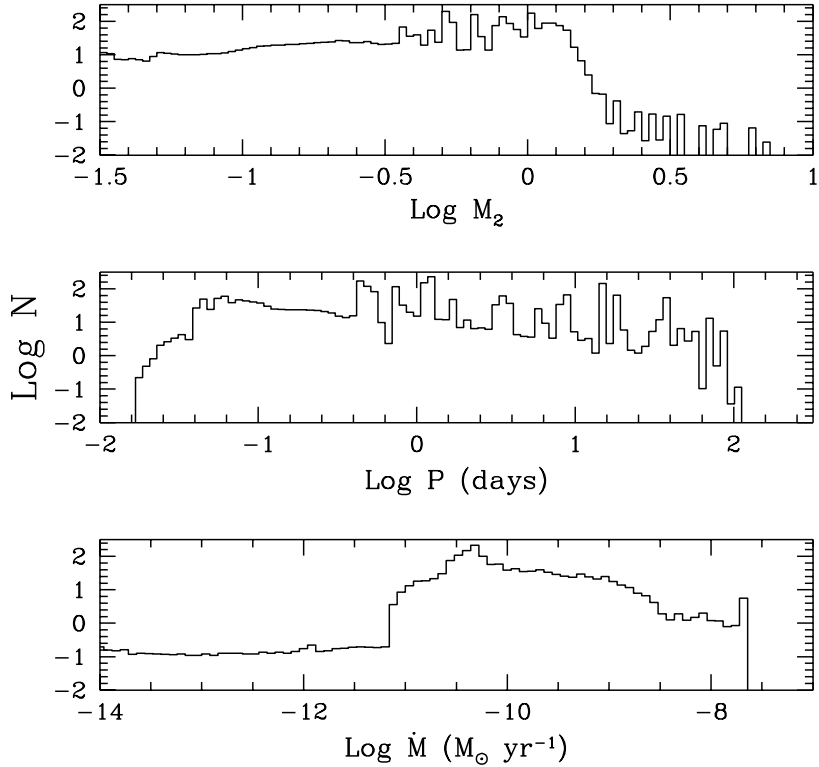,height=4.45truein,angle=0}}\hfill
\caption{The distributions of secondary mass (top panel), orbital period
(middle panel), and neutron-star mass-accretion rate at the current epoch
for our standard reference model. The jagged structure of the
histograms is a consequence of the discrete nature of the grid of 
binary sequences.}
\end{figure}

For our standard reference model, we pick an initial orbital period
distribution that is logarithmically flat, adopt a Miller-Scalo
initial mass function for the primary (Miller \& Scalo 1979) and a
flat distribution for the mass-ratio distribution. For the
supernova-kick distribution we choose the Maxwellian distribution with
$\sigma=190\,$km$\,$s$^{-1}$ from Hansen \& Phinney (1997). We plan to 
investigate the effects of varying these assumptions as part
of our study in the future. To determine the orbital parameters for 
post-common-envelope systems, we use binding energies for the common
envelope obtained from realistic stellar calculations with or without the 
inclusion of the ionization energy (Han et al.\ 1995; Dewi \& Tauris 2000).

\subsection{Preliminary results}

In Figures~6 and 7 we present some of the early results of our population
synthesis simulations for our standard set of assumptions. Figure~6 displays the
distributions of the secondary mass, orbital period and system space
velocity at the beginning of the X-ray binary phase. The mass
distribution is dominated by intermediate-mass systems; the more
massive systems (above $\sim 4\,M_{\odot}$) will experience (delayed)
dynamical mass transfer and will not contribute significantly to the
population of X-ray active systems. At the other end of the mass
spectrum, we find very few low-mass, CV-like systems.\par

Figure 7 in contrast shows the distributions of the secondary mass, orbital
period and neutron-star mass-accretion rate during the X-ray binary
phase at the current epoch. Now the mass distribution is dominated 
by relatively low-mass
systems, and there are hardly any systems above $\sim 2\,M_{\odot}$.
The reason is simply that the intermediate-mass systems lose most of
their mass rapidly in the initial super-Eddington phase and spend most 
of their lives at relatively low masses (see Figs~2 and 3). The orbital-period
distribution shows no period gap and extends to very low periods. Finally, the
luminosity distribution displays a fairly strong peak around $5\times 10^{-11}\,
M_{\odot}\,\hbox{yr}^{-1}$ and has a sharp cut-off at $\sim 10^{-11}\,
M_{\odot}\,\hbox{yr}^{-1}$.\par

While these distributions show many of the characteristics of the
observed distributions of ``LMXBs'', in fact more so than a model that
only includes CV-like systems, there are still some fairly obvious
discrepancies. First, there are too many short-period systems to be
consistent with the observed period distribution (e.g., Ritter \& Kolb 1998), 
although we note that, in this preliminary calculation, we did not 
consider whether systems are transients or not. Second, while the 
distribution of mass-accretion rate (and hence X-ray luminosity) has a 
sharp cut-off at $\sim 10^{-11}\,M_{\odot}\,\hbox{yr}^{-1}$ 
--- as is desirable, the peak in the distribution is 
probably too low by about an order of magnitude.

\section{Outlook}

The binary-population synthesis results presented in this contribution 
represent only a first step in our investigation. By defining a standard
reference model, we can quantitatively identify the problems with
this model (e.g., concerning the period and luminosity distributions
of LMXBs/IMXBs, the properties of binary millisecond pulsars, etc.).
Having identified the problems, we can then examine possible solutions.
Some of the modifications we plan to consider are cyclical mass transfer 
due to external irradiation, interrupted mass transfer and the role of the 
pulsar turn-on. To first order, we intend to implement these using a 
perturbation-style approach where we use the available grid of binary 
sequences to define the secular evolution.\par
Once we have developed a better model, we can then extend the present
calculations with the improved model on a much finer grid. With the
improvements in computing power and implementing a `minimalistic' version
of our binary stellar-evolution code, we estimate that even with present-day
workstations we should be able to calculate one whole binary sequence
in about 2 minutes. An attractive alternative is to use the `Eggleton' 
supercluster at the Lawrence Livermore Laboratory whose potential was
impressively demonstrated at this meeting (Eggleton 2001). 
The ultimate goal, of course, is to obtain a realistic modelling of the 
X-ray binary population, a goal that now appears achievable within
the next few years.



\end{document}